\documentclass[sigconf, 10pt, noacm]{acmart}

\AtBeginDocument{%
  }

\renewcommand\footnotetextcopyrightpermission[1]{}

\newcommand\DISCO{\textit{DISCO}}
\newcommand\DIPP{\textit{DIPP}}
\newcommand\proc{\textit{DISH}}

\usepackage{hyperref}
\usepackage{cleveref}
\usepackage{subcaption}

\usepackage{listings}
\usepackage{xcolor}

\definecolor{codegreen}{rgb}{0,0.6,0}
\definecolor{codegray}{rgb}{0.5,0.5,0.5}
\definecolor{codepurple}{rgb}{0.58,0,0.82}
\definecolor{backcolour}{rgb}{0.95,0.95,0.92}

\lstdefinestyle{mystyle}{
    commentstyle=\color{codegreen},
    numberstyle=\tiny\color{codegray},
    stringstyle=\color{codepurple},
    basicstyle=\ttfamily\footnotesize,
    breakatwhitespace=false,         
    breaklines=true,                 
    captionpos=b,                    
    keepspaces=true,                 
    numbers=left,                    
    numbersep=5pt,                  
    showspaces=false,                
    showstringspaces=false,
    showtabs=false,                  
    tabsize=2
}

\usepackage{array}
\usepackage{ragged2e}
\usepackage{multirow}
\newcolumntype{P}[1]{>{\centering\arraybackslash}p{#1}}
\newcolumntype{M}[1]{>{\centering\arraybackslash}m{#1}}
\newcolumntype{R}[1]{>{\RaggedLeft\arraybackslash}p{#1}}

\begin{document}

\title{Adaptive and Robust Image Processing on CubeSats}

\author{Robert Bayer}
\email{roba@itu.dk}
\orcid{0000-0002-1052-3855}
\affiliation{%
  \institution{IT University of Copenhagen}
  \city{Copenhagen}
  \country{Denmark}
}

\author{Julian Priest}
\affiliation{%
  \institution{IT University of Copenhagen}
  \city{Copenhagen}
  \country{Denmark}
}

\author{Daniel Kjellberg}
\affiliation{%
  \institution{IT University of Copenhagen}
  \city{Copenhagen}
  \country{Denmark}
}

\author{Jeppe Lindhard}
\affiliation{%
  \institution{IT University of Copenhagen}
  \city{Copenhagen}
  \country{Denmark}
}

\author{Nikolaj Sørenesen}
\affiliation{%
  \institution{IT University of Copenhagen}
  \city{Copenhagen}
  \country{Denmark}
}

\author{Nicolaj Valsted}
\affiliation{%
  \institution{IT University of Copenhagen}
  \city{Copenhagen}
  \country{Denmark}
}

\author{Ívar Óli}
\affiliation{%
  \institution{IT University of Copenhagen}
  \city{Copenhagen}
  \country{Denmark}
}

\author{Pınar Tözün}
\affiliation{%
  \institution{IT University of Copenhagen}
  \city{Copenhagen}
  \country{Denmark}
}

\renewcommand{\shortauthors}{Bayer et al.}

\fancyhead{}

\begin{abstract}
  CubeSats offer a low-cost platform for space research, particularly for Earth observation.
  However, their resource-constrained nature and being in space, 
  challenge the flexibility and complexity of the deployed image processing pipelines and their orchestration. 
%
  This paper introduces two novel systems, \DIPP~ and \proc, to address these challenges.
  \DIPP~ is a modular and configurable image processing pipeline framework 
  that allows for adaptability to changing mission goals even after deployment, while preserving robustness.
  \proc~ is a domain-specific language (DSL) and runtime system designed 
  to schedule complex imaging workloads 
  on low-power and memory-constrained 
  processors.

  Our experiments demonstrate that \DIPP's decomposition of the processing pipelines adds negligible overhead,
  while significantly reducing the network requirements of updating pipelines
  and being robust against erroneous module uploads.
%
  Furthermore, we compare \proc~ to Lua, a general purpose scripting language, and demonstrate its comparable expressiveness and 
  lower memory requirement.
  
\end{abstract}




\received{20 February 2007}
\received[revised]{12 March 2009}
\received[accepted]{5 June 2009}

\maketitle

\section{Introduction}\label{sec:introduction}

Nanosatellites, especially CubeSats \cite{disco2, estcube, scientific_cubesat, ardusat}, have democratized access to space, enabling more organizations to participate in space-related research, such as Earth observation 
\cite{AksoyVLDB22, denbyOrbitalEdgeComputing2020, eagleeye}.
However, the limited resources of CubeSats
present challenges for applications deployed on CubeSats. 
First, the limited network bandwidth necessitates more on board data processing,
while bounding the updatability
of the deployed programs,
to reduce the downlink and uplink network traffic, respectively.
Second, the limited memory restricts the complex orchestration of these programs.

\textbf{Flexibility.}
Existing image processing pipelines deployed on CubeSats
are often inflexible, relying on fixed predefined processing stages that cannot be easily modified once the satellite is in orbit \cite{scientific_cubesat}. 
This inflexibility is problematic for missions where scientific goals or customer requirements change frequently, requiring changes to the image processing pipelines.
The constrained and often asymmetrical network bandwidth typical of CubeSats further amplify this challenge, making uploading new pipeline binaries a significant bottleneck.
The rising demand for offloading machine learning to satellites, leads to even higher pressure on the uplink due to more frequent updates to models.
Despite these limitations and challenges, integrating machine learning on CubeSats is essential for enabling new use cases, such as onboard event detection and image filtering. 
Performing these tasks on board can lead to a more efficient use of the limited bandwidth, downlinking only the most relevant data \cite{kodan}.

\textbf{Robustness.}
The main reason behind the conventional approach of inflexible pipelines on satellites is avoiding mishaps due to frequent updates.
CubeSats are often utilized as test platforms for new technologies due to their low deployment costs.
This approach allows for experimentation with novel technologies where mission failure poses a lower risk.
Nonetheless, robustness and fault-tolerance are of utmost importance for all deployments, where physical intervention is impossible \cite{SatHealthMonitor}.

\textbf{Orchestration.}
Emerging Earth observation applications also require complex and flexible task orchestration.
However, existing 
approaches often lack a memory and bandwidth efficient way of providing expressiveness and adaptivity required for missions involving dependencies between observations, geofencing, or precise attitude control, such as those required for photogrammetry. 

To address the challenges of image processing on CubeSats,
this paper presents \DIPP~ and \proc.
\DIPP~ (\DISCO~Image Processing Pipeline) is a modular and configurable approach to image processing pipelines that allows adaptability to changing mission goals, while preserving the robustness of the system.
\proc~ (DIsco ScHeduler) is a DSL and runtime for Earth observation, enabling users to schedule complex workloads and subsequent processing of the imagery, while running on low-power real-time processors with limited memory.

More specifically, the contributions of the paper are:

    \textbf{\DIPP: Modular, reconfigurable, and fault-tolerant image processing framework for CubeSats.} \Cref{sec:dipp} introduces a new framework for image processing that decomposes the execution of image processing pipelines into \textit{modules}, which can be added, reordered, replaced, or removed with a set of parameters specific to the pipeline,
    while allowing module reuse across pipelines. 
    We evaluate \DIPP~using a representative image processing pipeline, composed of preprocessing, deep learning inference, and postprocessing modules.
    The results demonstrate that \DIPP~adds negligible overhead to end-to-end pipeline execution, while almost halving the data to be uplinked if one wants 
    to update functionality over time.
    Furthermore, any errors or failures that occur within a module are contained in the module.

    \textbf{\proc: Expressive DSL and runtime for complex observation scheduling on CubeSats.} \Cref{sec:proc} introduces a new DSL and runtime system.
    The DSL combines the low resource requirements of event-driven programming and the expressiveness of general-purpose scripting.
    The runtime polls the configuration and telemetry of other components in the satellite, such as the battery level and power generation, for power-aware scheduling and orchestration of flight plans.
    We illustrate that \proc~brings the expressiveness of a general-purpose scripting language, such as Lua, onto a hardware platform suitable for CubeSats, despite Lua being unable to fit on such memory-constrained platforms.

The rest of this paper is organized as follows. \Cref{sec:background} presents the educational satellite that this work targets and the related work.
\Cref{sec:evaluation} evaluates the performance of \DIPP~and \proc, while \Cref{sec:discussion} discusses trade-offs of their design choices and the future work directions.
Finally, \Cref{sec:conclusion} concludes the paper.

\section{Background}\label{sec:background}

To better illustrate the target satellite use case for this work,
\Cref{sec:background:disco}
introduces the \DISCO~project.
Then, 
\Cref{sec:background:relatedwork}
surveys related work, 
with a focus on 
in-orbit processing and observation scheduling.



\subsection{\DISCO-2}\label{sec:background:disco}

\textbf{Goal \& Requirements.} \DISCO~is an educational satellite program, where students are given the opportunity to design and operate CubeSats, a class of nanosatellites with a form factor of 10cm cubes per unit and the possibility to stack multiple units to create a larger satellite. It is a collaboration between three universities and the Polar Research Organization. 

\DISCO-2 is the second satellite to be launched in the project. 
It is a 3U CubeSat carrying an Earth observation \textit{payload}, including two RGB
and a single IR (infrared) camera. 
Besides capturing images, the payload will perform in-orbit image analysis to increase the value of the downlinked data.
While the payload will initially be tasked with a simple cloud cover detection, the long-term goal is to provide a platform for research,
such as photogrammetry of glaciers in the Arctic.

Because of \DISCO-2's asymmetrical bandwidth, achieving 1mbit/s downlink and 9.6kbit/s uplink, it is crucial to utilize the bandwidth efficiently. 
This constraint and the need to support  frequently changing workloads, 
require a 
reconfigurable image processing pipeline. 
Furthermore, while all the tasks submitted for deployment on the satellite will be thoroughly tested for errors on the satellite's replica on Earth, the image processing pipeline needs to be highly fault-tolerant against errors not caught during the tests.

Earth observation tasks also need complex orchestration, such as attitude control or precise timing.
To enable deployment of such tasks, we need a way to express their logic in a compact way that can run on a memory-limited core.

\begin{figure}[t]
\centering
\includegraphics[width=\linewidth]{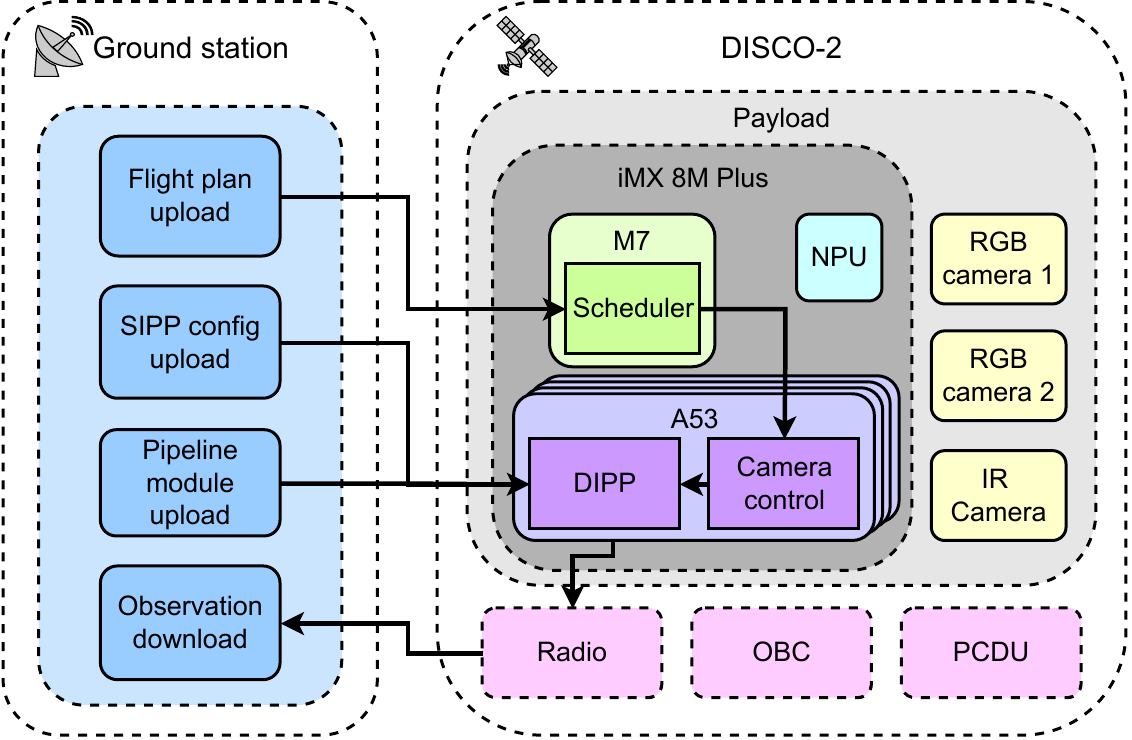}
\caption{Simplified architecture diagram of \DISCO-2, focusing on the satellite's payload and highlighting interactions between the satellite and a ground station.}
\label{fig:sat-arch}
\end{figure}

\textbf{Architecture.}
\Cref{fig:sat-arch} shows a simplified architecture of \DISCO-2 with specific focus on the payload and the interactions of the components on board and a ground station.

Communication between the satellite and ground station/s and between the components of the satellite (Radio, on-board computer (OBC), power unit (PDU), cameras, M7, etc.) uses the CubeSat Space Protocol (CSP), a popular lightweight networking protocol for nanosatellites modeled on TCP/IP.
CSP exposes each component as a node on the network for targeted communication.
Each node on the network also exposes a parameter table, allowing any node on the network to read and change values in this table containing node's configuration and telemetry.

To enable running computationally demanding processing in a power-efficient way,
the payload contains the iMX 8M Plus processor \cite{imx8, ReachingEdgeEdge}.
This chip provides a heterogeneous multiprocessing platform, which includes a single low-power ARM Cortex-M7 real-time core running FreeRTOS, four powerful ARM Cortex-A53 application cores running custom Linux-based operating system, and an ARM NPU machine learning inference accelerator (see \Cref{tab:imx8-spec}). 
The M7 core is dedicated to the observation scheduler and manager (\proc), 
the four A53 cores are dedicated to running the camera controller and the image processing pipelines (\DIPP), and 
the NPU does not run anything by default, but can be employed by a pipeline module to accelerate machine learning inference.
While the chip also contains a GPU, we have not found a workload where it would outperform the A53 cores or NPU, and therefore do not utilize it. 

\begin{table}[t]
\centering
\begin{tabular}{@{}lR{8em}R{9.5em}@{}}
\toprule
Cores & Cortex-M7 & 4x Cortex-A53 \\
\midrule
Cache & 32KB I\$, 32KB D\$ & 32KB I\$, 32KB D\$, 512KB L2\$ \\
Memory & 128KB ITCM, 128KB DTCM & 2GB LPDDR4 \\
Storage & 2KB EEPROM & 8GB eMMC \\
\midrule
Accelerator & \multicolumn{2}{c}{2.25 TOPS NPU} \\
\bottomrule
\end{tabular}
\caption{Specs of iMX 8M Plus processor components.}
\label{tab:imx8-spec}
\end{table}


\textbf{Operation.} Observations are scheduled from a ground station, from where operators upload a flight plan\footnote{A sequence of commands to be performed by the satellite.} to the scheduling system on the satellite.
This system admits these plans, and schedules them for execution, polling other systems on the satellite for values if flight plans require external inputs.
Once the system is ready to perform an observation, it wakes up the application cores and executes image capture through a camera control system.
The camera control captures images and stores them in memory to be accessed by the image processing pipelines. 
Execution of these pipelines can be triggered by the scheduler at a later stage in case of heavy workloads or power-aware scheduling, or the pipeline can wait for incoming images in polling mode and execute immediately after capture.
The processed output is then offloaded to a persistent buffer, from where the operators can retrieve them.
Lastly, the configurations of pipelines and their modules are uploaded directly to \DIPP~together with possible missing modules.

\subsection{Related Work}\label{sec:background:relatedwork}


\textbf{In-orbit image processing.} 
One of the biggest challenges of nanosatellites is their highly limited downlink bandwidth. 
While the operators can increase the downlink budget linearly by renting ground stations around the world, this goes against one of the principles of nanosatellites, their cost-efficiency \cite{CommunitydrivenApproachDemocratize2021}. 
An alternative is in-orbit processing of images before sending them back to Earth.
An early example of onboard data processing using machine learning is the $\phi$-Sat-1 deployed by the European Space Agency \cite{phisat}. 
This satellite employed deep convolutional neural networks to recognize cloud-covered images, which has since become one of the most common use cases \cite{ReachingEdgeEdge}, leading to up to $\sim 67\%$ reduction in number of images sent to Earth \cite{modis},
with more complex processing providing even higher savings \cite{kodan, FinalFrontierDeep}.

Besides machine learning, processing of satellite imagery often involves 
operations such as compression, cropping, etc.
The HYPSO project \cite{hypso1, hypso1-ocean, hypso-pipeline} 
features predefined image processing pipelines that can be reconfigured after launch.
However the pipelines are not fully decomposed and updates therefore require uplink of the entire pipeline binary.
Another example is the Onboard Data Processing Software deployed in the HyperScout platform by the European Space Agency \cite{hyperscout}. 
This work focuses on real-time processing by utilizing job templates, 
which are passed to parallel worker threads running predefined sets of pipeline modules.
\DIPP~extends on HYPSO and HyperScout with higher degree of flexibility via the ability to incrementally upload pipeline modules and extend the capabilities of the pipelines beyond the pre-defined modules the satellite is launched with.

\textbf{Module isolation.}
Not allowing
new pipeline modules after deployment of the satellite provides a higher degree of reliability.
To allow flexibility without breaking reliability,
\DIPP~ isolates the pipeline modules from one another and from the other systems on the satellite
(\Cref{sec:dipp:exec}). 
\DIPP~employs process isolation and monitoring inspired by the supervisor design pattern presented by Erlang \cite{erlang} and Hao \cite{haoElixir}.

\textbf{Remote orchestration of embedded devices.}
Ad-hoc offloading of tasks to satellites \cite{SatServiceOrchestration} is becoming a reality on larger satellites.
However, the resource-constrained nature of nanosatellites makes ad-hoc offloading of general computing tasks challenging, due to their limited network bandwidth and memory.
While many works embed the Lua scripting language \cite{lua} in programs to extend functionality at runtime, Lua was not developed for embedded systems and its requirements exceed resources available on nanosatellites, unless stripped off many of its features \cite{hempelLuaPort}.

The Canadian Space Agency has explored the offloading
of flight plans to satellites
\cite{desouzaFlightSoftwareDevelopment2022}. 
They implemented a timing and queue-based flight plan system integrated into CSP.
However, this implementation lacks the expressiveness for defining complex flight plans. 
\proc~closes this gap by providing a more expressive framework
(loops, recursion, dependencies across concurrent flight plans ...),
which can be scheduled and dispatched by low-power processors on nanosatellites.


\begin{figure*}[ht]
\centering
\includegraphics[width=\linewidth]{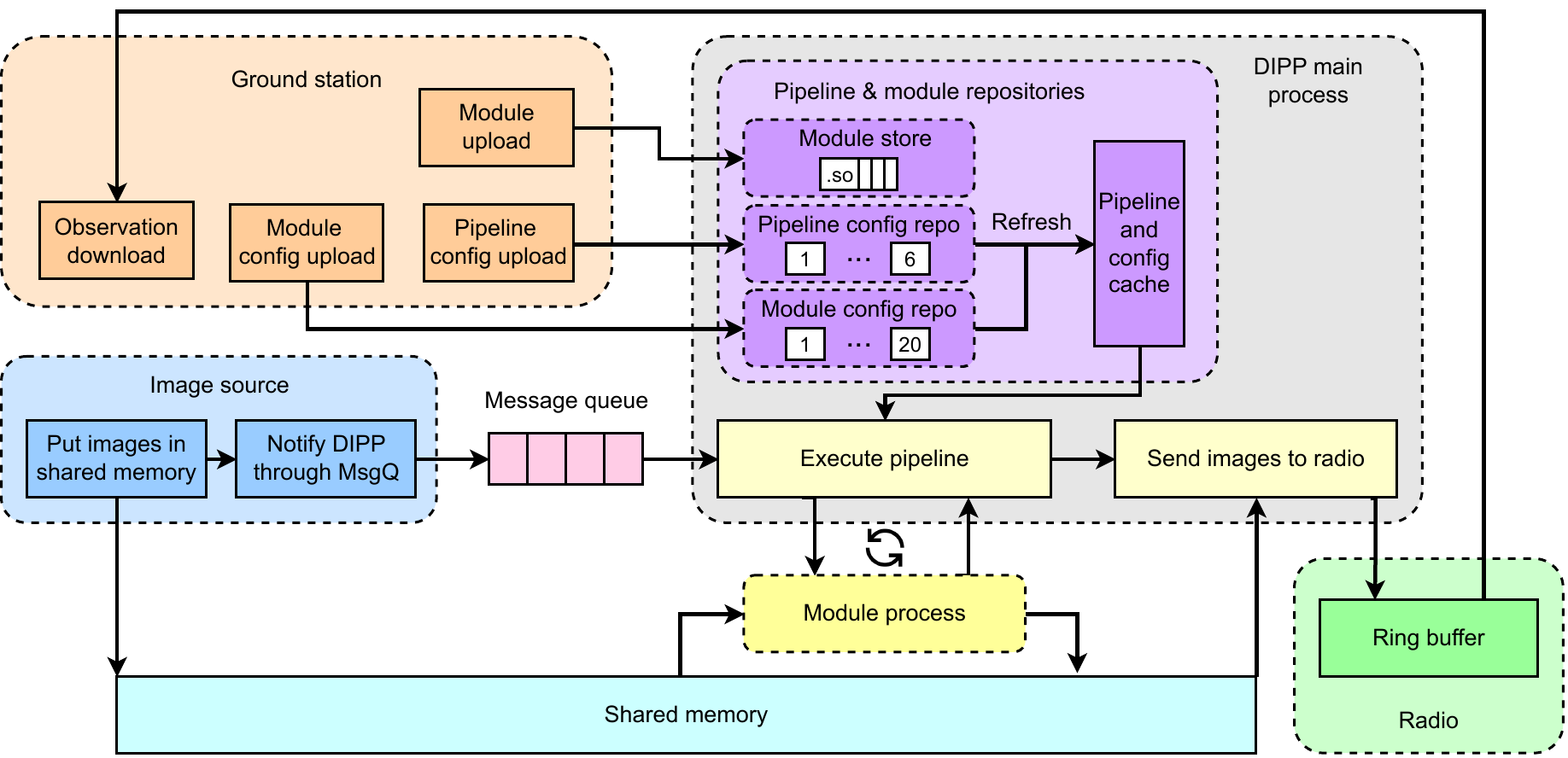}
\caption{System architecture diagram of \DIPP. \textit{Image source} places images in \textit{Shared memory} and notifies \DIPP~via a \textit{Message queue}. \DIPP~loads cached modules and configurations, executes them, and sends processed images to the \textit{Radio}. The \textit{Ground station} manages \DIPP~configuration uploads, module uploads, and observation downloads. When new configurations are uploaded, pipeline and configuration \textit{cache} is invalidated and rebuilt.}
\label{fig:dipp-arch}
\end{figure*}

\section{\DIPP}\label{sec:dipp}

\DIPP~is comprised of services, components, and communication channels that together form the framework for configuration and execution of the image processing pipelines on the satellite.
\Cref{fig:dipp-arch} shows the system architecture of \DIPP~and its interaction with other nodes on the satellite's network, including the image source, ground station, and radio.

\DIPP~receives images from the image source, which in the case of \DISCO-2 is the camera control software operating the three cameras on board of the satellite.
The image transfer relies on two channels (detailed in \Cref{sec:dipp:comm}):
(1) metadata about the image batch is communicated through a message queue and
(2) the raw image data is shared between the processes through shared memory to accommodate the larger size.
The camera controller can send a single image or batch similar images taken in succession. 

\DIPP~stores a repository of image processing modules and configurations. 
Each module represents a single data processing function with a single entry point, accepting a batch of images as input and returning a processed batch as output. 
They are compiled as independent shared object files (.so) that can be uploaded to the satellite and integrated into pipelines even after launch.
Each module can either completely encapsulate its functionality by statically linking its dependencies or rely on dynamically linked libraries already present on the satellite avoiding the need to upload large libraries repeatedly.
To ensure compatibility and ease of setup, users are provided with module templates, which can be adjusted to perform the custom functionality.

Two types of configurations govern the pipeline's behavior (detailed in \Cref{sec:dipp:conf}): (1) pipeline structure configurations, which determine the sequence of modules to be executed, and (2) module parameter configurations, which provide runtime arguments for modifying module behavior. 
These configurations are defined in YAML and uploaded from the ground station using custom commands 
throught CSP.

When a new image batch arrives in the message queue for processing, \DIPP~dynamically loads and executes the appropriate modules based on the configurations
(detailed in \Cref{sec:dipp:exec}). 
The subsequent pipeline calls re-use these loaded modules.
Changes to the configurations of pipelines or modules, or module uploads invalidate these cached pipelines, triggering reloading based on the updated configurations.
Each module runs in an isolated child process, preventing errors in one module from affecting the entire pipeline.
Furthermore, the main process monitors the execution of the modules and implements recovery strategies in case of failure. 
For efficiency, image data remains in shared memory, with modules directly accessing and modifying it. 
Only metadata is passed between modules and the main \DIPP~process.

After processing, the resulting data is transferred to the on-board radio module for downlink to the ground station. 
Storing data on the radio module prevents the need for the power-hungry application cores to be awake during retrieval of the observations. 
The radio module uses a ring buffer managed through the CSP parameter table interface, overwriting old observations to prevent running out of storage.

\subsection{Communication}\label{sec:dipp:comm}

Each software component of the satellite payload (\Cref{fig:sat-arch}) runs in a separate process with its own memory space. 
Communication between the processes has to therefore be facilitated using inter-process communication (IPC).
This is primarily a challenge for
communicating the new image batches
between the camera controller and \DIPP.
This communication should adhere to some requirements: 
\begin{list}{\labelitemi}{\leftmargin=1.5em}
    \item \textbf{Asynchronous Operation.} The camera controller must not be blocked waiting for \DIPP~to receive the data, allowing the camera to work on the next batch independently.
    \item \textbf{Large Data Size Support.} Large image batches must be accommodated without imposing size restrictions.
    \item \textbf{Metadata Transfer.} Along with the raw image data, metadata of the batch must be transferred efficiently.
\end{list}

\subsubsection{Transfer of Images} \label{sec:dipp:comm:image}

Based on the requirements, we explored three IPC techniques:
shared memory and two message passing techniques, pipes and message queues.

\textbf{\textit{Shared memory}}
allows multiple processes on the same machine to access a shared region of the physical memory and its size is bound only by the size of the physical memory.
Processes can directly read from and write to the shared memory.
Shared memory is also asynchronous in nature, allowing a process to transfer data while the receiving process is inactive or yet to be created.
The largest drawback is the lack of implicit concurrency control, leading to risks of race conditions.
Furthermore, the processes need to agree on an identifier (ID) for the shared memory segment.
This ID is used when creating the shared memory segment and is needed by other processes to attach to that segment.
If only a few fixed shared memory segments are needed, the processes can agree on a set of static IDs, requiring no further communication.
The problem arises when multiple shared memory segments are dynamically created in a system, thus requiring another IPC technique to communicate these IDs.

\textbf{\textit{Pipe}} is a unidirectional message passing technique implemented as a first-in, first-out (FIFO) queue.
Pipes allow for communication between two or more processes using two operations: \textit{send} and \textit{receive}, providing implicit concurrency control.
However,
they provide limited capacity.
Since Linux 2.6.11, pipes are limited to 16 pages (16 × 4096 bytes), meaning efficient transfer of data larger than the pipe's capacity requires synchronous communication.

\textbf{\textit{Message queues}} offer a bidirectional communication channel and are not bound by the FIFO order, unlike \textit{pipes}.
However, similar to \textit{pipes}, they are limited in the amount of data they can store without being emptied,
and similar to \textit{shared memory}, processes need to agree on identifiers.


While all three techniques provide asynchronous data access, this property is not truly preserved for moving large amount of image data in the case of pipes and message queues,
leaving shared memory as the only option for communicating the image data.


\subsubsection{Transfer of Image Batch Metadata} \label{sec:dipp:comm:metadata}

The choice of using shared memory for the images comes with the requirement of managing IDs for the shared memory segments.
There are two options for this,
as \Cref{sec:dipp:comm:image} hinted.

Creating \textbf{\textit{a large memory segment}} with static segment IDs would eliminate the need to dynamically communicate the IDs and reduce overhead of creating and managing multiple shared memory segments.
However, this requires a complex memory manager to coordinate data access. 
Processes need to synchronize their reads and writes carefully, potentially leading to performance bottlenecks.
Furthermore, if a pipeline module resizes the images,
this would either overwrite other data in the shared segment or require padding of the input data batch to allow for the possible expansion.



Creating \textbf{\textit{a separate memory segment for each batch}}
would eliminate the need for synchronization.
Furthermore, \DIPP~modules can resize images without affecting other data by simply resizing their associated shared memory segments.
However, this increases overhead by creating separate segments and requires an additional IPC mechanism to communicate the segment IDs between the camera controller and \DIPP.
Overall, for our target use case, we found the overhead of creating image segments to be negligible
(\Cref{sec:evaluation}).
To communicate the shared memory segment IDs, we make them part of the image batch metadata.

We use message queues to transfer the image batch metadata.
As opposed to pipes, message queues are made to share structured data, giving us a structured way to transfer the batch metadata.
Additionally, message queues provide the option for out-of-order retrieval and message type identification, which one can leverage for image batch prioritization.

\begin{figure}[t]
\centering
\includegraphics[width=\linewidth]{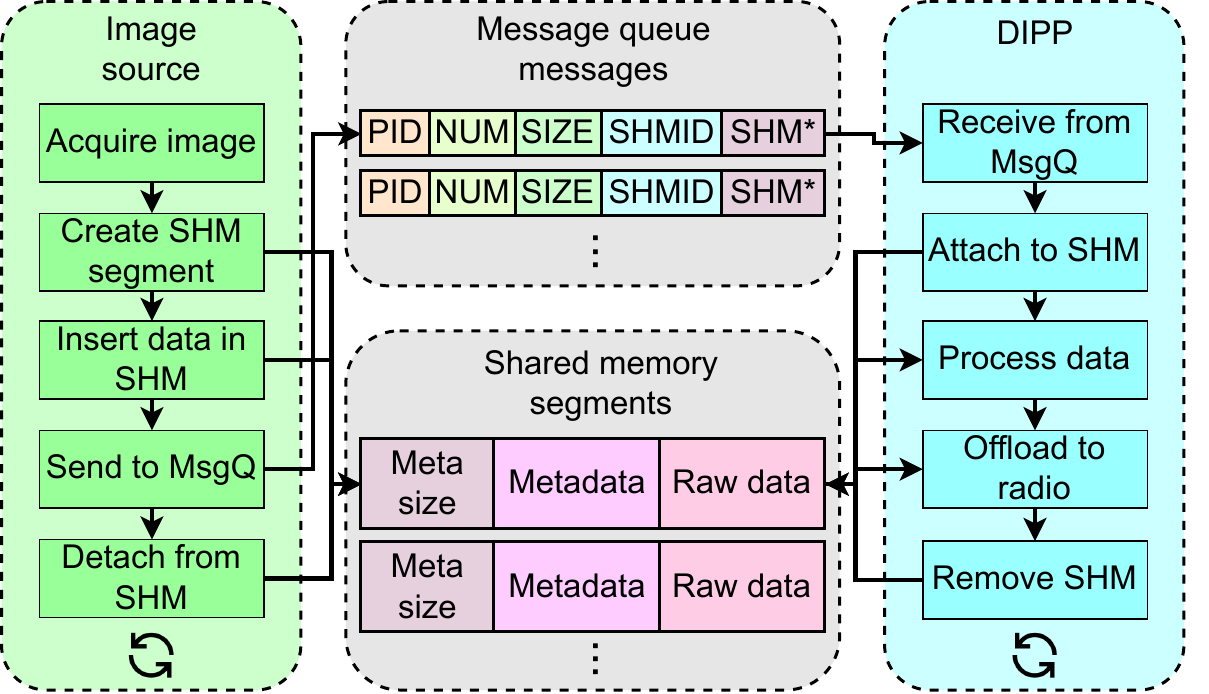}
\caption{Data transfers between the camera controller and \DIPP~using a message queue and shared memory. This design provides asynchronous data transfer, where metadata for the shared memory segment arrives through the message queue and image data resides in shared memory to avoid size limitations.}
\label{fig:dipp-comm}
\end{figure}

\subsubsection{The final design} \label{sec:dipp:comm:comm}

Based on the analysis in Sections~\ref{sec:dipp:comm:image} and \ref{sec:dipp:comm:metadata},
the proposed solution for communication between the image source and \DIPP~ combines shared memory and message queues, as shown in \Cref{fig:dipp-comm}.
Image source acquires a batch of images and creates a shared memory segment.
It then copies the raw image data to the memory segment and constructs a message containing metadata about the image batch.
This metadata contains a pipeline ID (PID) to be used for processing the image batch, the number of images in the batch (NUM), the total size of the image batch in bytes (SIZE), the unique identifier of the shared memory segment containing the raw image data (SHMID), and a pointer to the starting address of the shared memory segment.
It then sends this message to the message queue, notifying \DIPP~that the batch is available for processing. 
Afterward, it can detach from the shared memory segment.

Meanwhile, \DIPP~can wait for the arrival of the next message on the queue or be triggered manually by the scheduler (\proc). 
Once \DIPP~retrieves a message from the message queue, it extracts the shared memory segment ID and the rest of the image batch metadata.
\DIPP~ uses the segment ID to attach to the segment of the raw image data.
Each raw image in a batch is structured as a contiguous memory starting with the size of the image metadata, followed by the serialized image metadata, and the raw image bytes.
The image metadata contains size of the image in bytes, indicating how many bytes to read past the metadata, the image size in pixels, color depth, timestamp, and the associated camera ID. 
This metadata is serialized using Protobuf \cite{protobuf}.
In addition to these default parameters, users can implicitly extend the metadata with custom parameters with varying data types.
\DIPP~then processes these images (elaborated on in \Cref{sec:dipp:exec}), offloads them to a persistent buffer, and finally removes the memory segment.

\subsection{Configuration}\label{sec:dipp:conf}
The adaptability of \DIPP~to changing mission requirements is achieved through upload of new modules, and pipeline and module configurations.
While new modules are uplinked as shared object files using CSP's data transfer protocol, the configuration of pipelines and their modules is handled through custom commands in CSP shell (CSH).

\begin{figure}[t]
\centering
\includegraphics[width=\linewidth]{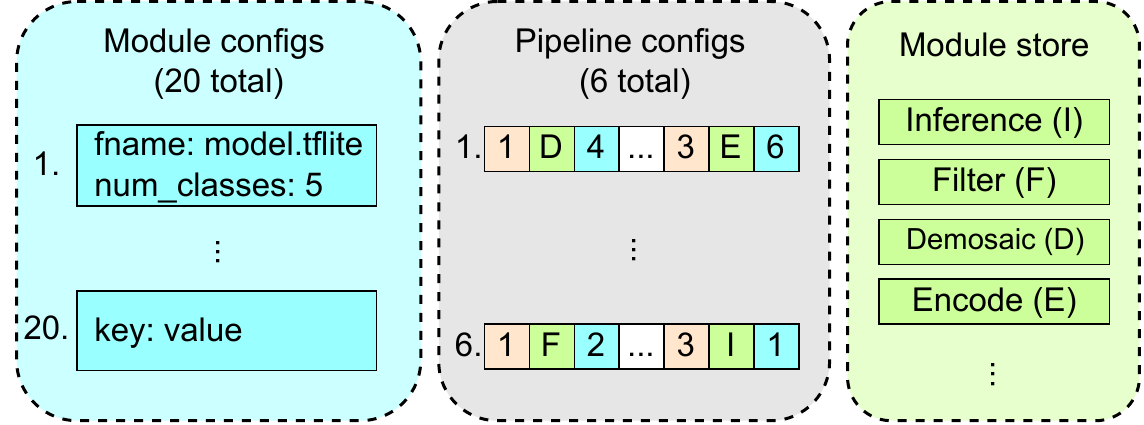}
\caption{Example \DIPP~configuration. The left box lists 20 module parameters configuring individual modules. The middle box defines 6 processing pipelines, each specifying a sequence of modules. Each module definition is comprised of three blocks: first (orange) indicates a module’s sequence within the pipeline, second (green) refers to a module in the module store, and third (blue) refers to the module configuration ID to be associated with (0 means no association). The right box lists the modules available as shared object files.}
\label{fig:dipp-conf}
\end{figure}

One of the CSH commands is for configuring pipelines. 
Users provide YAML files representing the pipeline configurations.
The pipeline configuration repository provides multiple slots.
This allows users to specify multiple pipeline configurations specific to each camera.
The pipeline ID (PID), introduced in \Cref{sec:dipp:comm:comm}, represents one of these configurations to be executed.
As \Cref{fig:dipp-conf} shows, pipeline configurations contain three parameters for each module: (1) its order within the pipeline, (2) its name, and (3) the ID of the configuration to be applied to this module in this pipeline.

There are also slots for module configurations. 
Multiple pipelines can use the same module without being restricted by a single configuration.
For example, an inference module can be invoked with different models and input by different pipelines.
The module configuration provides keyword arguments, which modules can retrieve at runtime.
This allows users to fine-tuning processing performance dynamically: 
Operators can balance processing speed, compression ratios, and image quality to meet mission-specific requirements.

\DIPP~ similarly provides a CSH command through which operators can update module configurations.
This command also accepts YAML files, which are serialized using Protobuf and compressed using Brotli \cite{brotli}. 
This ensures that the configurations are serialized and deserialized correctly and minimizes the amount of data to be transferred.
These compressed serialized representations are then uploaded into \DIPP's CSP parameter table which serves as the pipeline and module configuration repository.
The use of CSP parameter table provides flight tested means of ensuring correct upload and integration of these configurations.
CSP parameter table also provides callbacks on parameter changes, which are used to invalidate and reload the already loaded pipelines.

\subsection{Execution}\label{sec:dipp:exec}


The execution of pipelines starts with the main \DIPP~process retrieving message from the message queue (\Cref{fig:dipp-comm}),
including the shared memory segment ID, 
and attaching to the corresponding segment.
If this is the first execution of the corresponding pipeline,
it loads the pipeline in memory.
Otherwise, it uses the pre-loaded pipeline from a prior invocation and proceeds to execute it.
It loops through the modules in the pipeline and creates a child process for each module.
In the child process, \DIPP~ calls the module's entry point, supplying it with the image batch metadata and a pointer to start of the shared memory segment.
Once the module is executed, the child process communicates the output using an anonymous pipe or an error pipe. 
In case the module executed correctly, \DIPP~ proceeds to the next module.
Otherwise, if the error pipe was not empty or the child process crashed, \DIPP~ halts the execution of the pipeline and invalidates the cache.
In addition to errors and crashes, the parent process detects hanging processes and applies configurable timeouts to the modules, which are handled similarly.

In order to signal these errors, we developed a custom error code scheme. These are six digit integers where the first three digits are module or pipeline error codes, the fourth digit represents the pipeline ID, and the last two digits represent the module sequence in the pipeline.
The error codes correspond to errors with message queues, shared memory, loading of the shared object files, etc.
Additionally, a range of error codes is allocated to modules.
Each module can assign meaning to the error codes providing flexibility.
These error codes, together with the rest of telemetry are then visible in the \DIPP's parameter table for observability.

\section{\proc}\label{sec:proc}

\proc~aims at facilitating observations
by providing a framework for defining and executing flight plans, sequences of actions that control the satellite's behavior.
Before our custom solution, we explored two already available solutions.

\textbf{Event-driven programming.}
While simple and intuitive, this model, involving a programmable queue of events triggering actions, lacked the flexibility to express the complex, potentially evolving nature of flight plans. 
This approach would require constant code updates and re-uploads to accommodate new scenarios or scientific objectives, which is impractical for a satellite in orbit.

\textbf{General-purpose scripting languages.}
This would have offered considerable flexibility, but embedding a general-purpose scripting language, like Lua proved to be a significant challenge on our target device (\Cref{sec:background:disco}) or devices commonly used as on board computers on nanosatellites.
The Cortex-M7 core provides 128KB of instruction tightly-coupled memory (ITCM).
Lua alone requires 92.8KB of this limited memory, while this memory also needs to contain the ARM embedded toolchain, manufacturer's SDK, and the CSP library for communication.
These altogether led to 198.4KB required ITCM, significantly exceeding the available resources.
As mentioned in \Cref{sec:background:relatedwork}, some works managed to strip Lua down significantly, making it possible to deploy Lua on embedded devices.
However, this requires significant engineering effort.
Furthermore, this effort does not include integrating Lua with the CSP parameter tables to access data on other nodes of the satellite network, which is a highly desirable feature for flight planning.

\textbf{Custom solution.}
To avoid the cons and benefit from the pros of the above methods,
we create a custom solution, \proc, for scheduling observations.
\proc~provides a DSL and a runtime, with the aim of combining the flexibility of embedding a general-purpose scripting language with the lightweight nature of the event-driven programming model.
The library furthermore provides CSP client and server utilities for definition and subsequent deployment of flight plans.

\subsection{Procedures}\label{sec:proc:procedure}

\proc~ follows a simple register-based virtual machine model, where each procedure is a sequence of instructions that can be executed in a linear fashion.
The execution of a procedure can be modeled as a state machine, with each instruction causing the system to transition from one state to another. 
The state of the system can be represented as a tuple that includes the current instruction and the values of all parameter table entries across all nodes in the CSP network. 
However, since outside factors impact the value of the entries in parameter tables, the state transition system is not a pure function. 
Consequently, the execution of a procedure is not deterministic, and the system may transition to different states depending on the values of the parameters at the time of
execution, independent of the instruction being executed.

A procedure is represented as an array of instructions.
Each instruction can fall into three categories: (1) control flow, (2) arithmetic, and (3) memory operations.

There are four control flow instructions:
\begin{list}{\labelitemi}{\leftmargin=1.5em}
    \item \textit{proc block <param a> <op> <param b> [node]:} Pauses the execution of a procedure until a specified condition, involving parameters from the CSP network, is met. The \textit{[node]} represents the node where the parameters reside.
    \item \textit{proc ifelse <param a> <op> <param b> [node]:} Skips the next instruction if the condition is not met, and the following instruction if it is met.
    \item \textit{proc noop:} Performs no operation. This instruction is useful in else branches of the the ifelse instructions.
    \item \textit{proc call <procedure slot> [node]:} Inserts a call to another procedure. This enables the creation of more complex procedures by composing them from smaller units.
\end{list}

Also, there are unary and binary arithmetic operators:
\begin{list}{\labelitemi}{\leftmargin=1.5em}
    \item \textit{proc unop <param> <op> <result> [node]:} Applies a unary operator to a parameter and stores the result. The operator can be one of: $++$, $--$, $!$, $idt$, where $idt$ is the identity operator, useful for moving register values.
    \item \textit{proc binop <param a> <op> <param b> <result> [node]:} Applies a binary operator to the two parameters and stores the result. The operator can be one of: $+$, $-$, $*$, $/$, $\%$, $\ll$, $\gg$, $\&$, $|$, $\hat{}$.
\end{list}

Lastly, there is a memory instruction \textit{proc set <param> <value> [node]}, which sets the value of a parameter.

Users can define reusable procedures, which are compiled with \proc~ and available from the reserved procedure slots.
This minimizes the network requirements, by reusing common complex procedures such as interaction with Global Navigation Satellite System (GNSS) for geofencing, rather than including these in every flight plan upload.

Listing \ref{lst:proc:example} gives a a simple flight plan / procedure example, which waits until the GNSS time reaches a certain value, captures an image using the camera controller when it does, and processes this single-image batch using \DIPP.

\begin{lstlisting}[language=Python, caption=Example of procedure / flight plan, label=lst:proc:example]
proc new
# [START] Observation procedure 
# Observation start time
proc set p_uint32[0] 1744407764 $M7  
proc block gnss_time >= p_uint32[0] $M7
proc set wake_a53 1 $M7
# wait for the wake-up
proc block a53_status == 1 $M7
# Log the time of the observation
proc unop gnss_time idt p_uint32[1] $M7  
# Take image using camera 1
proc set capture_image 1 $CAM  
# Run image processing pipeline
proc set pipeline_run 1 $DIPP  
# Wait for image processing to finish
proc block pipeline_run == 0 $DIPP  
# Suspend A53 node
proc set suspend_a53 1 $A53  
# [END] Observation procedure
proc push 42 $M7
proc run 42 $M7
\end{lstlisting}

\subsection{Runtime}\label{sec:proc:runtime}

The runtime is responsible for interpreting the instructions (from \Cref{sec:proc:procedure}) and performing the corresponding operations on the state of the system. 
It also manages the control flow, procedure calls, and concurrent execution.

On a surface level, the runtime is a simple loop that iterates over the instructions in a procedure and executes them in a sequence. 
The state in \proc~ is largely encapsulated in the parameter table functionality of CSP, thus constraining the runtime’s responsibility to managing the control flow.
The \proc~runtime is available for both POSIX and FreeRTOS platforms, allowing for flexible deployment and portability in case of failure of the ARM Cortex-M7 real-time core.

\proc~ follows similar process isolation to \DIPP. 
When a new procedure is deployed for execution, \proc~ creates a new runtime child process, which provides isolation. 
This approach has multiple advantages including concurrent execution of procedures and higher fault-tolerance, as was highlighted in \Cref{sec:dipp:exec}.
Since complete process isolation is unachievable in FreeRTOS, where runtimes are dynamic tasks rather than processes, it does not provide the same level of fault-tolerance guarantees as POSIX platforms.

Inside a runtime, the procedures are executed by maintaining an instruction counter corresponding to the index of the instruction currently being executed in the procedure's instruction array.
When the \textit{block} instruction is executed, the runtime will periodically check the condition until it is met, while yielding CPU for execution of other procedures in between checks.
The arithmetic instructions
are delegated to the native C operators with an adapter layer that handles interaction with parameter tables.
Upon calling the \textit{call} instruction, a static analysis is performed, which builds a call graph.
This call graph is used to pre-fetch the dependent procedures.
Additionally, in case a \textit{call} instruction is a tail call, the runtime returns to the caller, updating the instruction counter and active procedure accordingly.
This optimization avoids nested recursion and memory overhead associated with running conventional loops.
Otherwise, the runtime simply nests procedure execution within the current one.



\subsection{Memory Allocation}\label{sec:proc:malloc}

Dynamic memory allocation is not desirable in embedded systems operating in critical environments, due to less predictable memory usage, memory fragmentation, and potential memory leaks.
While the isolated runtime process on POSIX systems provides robustness guarantees against most memory-related errors,
the same is not true on FreeRTOS.
We therefore provide the option of compiling  \proc~ with or without dynamic memory allocation.
This allows \proc~
to leverage the higher robustness guarantees on POSIX platforms, minimizing the amount of reserved memory,
while minimizing memory errors on FreeRTOS with statically allocated memory.
In case of FreeRTOS, indicating more constrained platforms,
\proc~will likely be the main application running on the processor and reserving large portion of memory statically therefore does not limit other applications.

\proc~ creates an abstraction layer providing platform-agnostic memory API to allocate, resize, and free blocks of memory.
On POSIX platforms, this API wraps the \textit{malloc}, \textit{realloc}, and \textit{free} functions.
FreeRTOS only provides \textit{pvPortMalloc} and \textit{pvPortFree}.
We therefore implement our own function for resizing the memory blocks.

\subsection{Serialization}\label{sec:proc:serialize}

To transmit the procedures from a ground station to the satellite, \proc~serializes the procedures sequentially, iterating over the procedure's instructions and serializing each one independently. 
The serialized CSP packet includes one byte for \proc-specific transmission flags such as request type (e.g., pushing or deleting a procedure to/from a slot), one byte for the procedure slot number, and the serialized procedure in the remainder of the packet.
The serialized procedure contains a byte representing the total number of instructions within the procedure followed by the serialized instructions.
Each instruction is composed of a two-byte CSP node number, one-byte instruction type, and variable number of bytes for instruction data.
For example, for a unary operation the instruction data would consist of one byte for each character in the operand parameter name and an additional byte for the null terminator, followed by a single byte representing the operator type, and finally the result parameter name, encoded in the same fashion as the operand.

\section{Evaluation}\label{sec:evaluation}

We now evaluate
\DIPP~(\Cref{sec:eval:dipp}) and \proc~(\Cref{sec:eval:proc})
on the Cortex-A53 and Cortex-M7 cores, respectively, of
the target hardware for \DISCO-2 (\Cref{tab:imx8-spec} \& \Cref{sec:background:disco}).

\subsection{\DIPP}\label{sec:eval:dipp}

While evaluating \DIPP,
our goal is to investigate 
its end-to-end performance for a complex image processing pipeline including deep learning inference (\Cref{sec:eval:dipp:e2e}),
the impact of its modular design (\Cref{sec:eval:dipp:overhead}),
its robustness against errors (\Cref{sec:eval:dipp:robustness}),
and its memory footprint (\Cref{sec:eval:dipp:scalability}).


\begin{figure*}[t]
\centering
\begin{subfigure}{.5\linewidth}
  \centering
  \includegraphics[width=\linewidth]{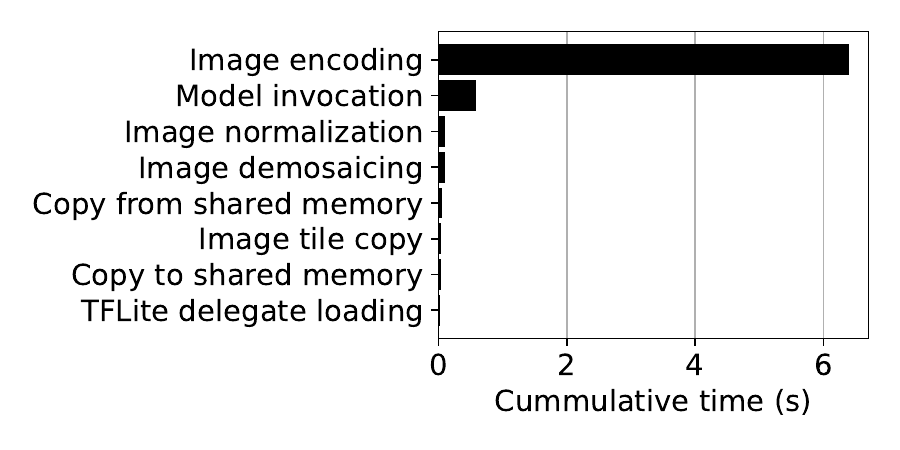}
  \caption{Decomposed execution of \DIPP.}
  \label{fig:dipp-breakdown-multi}
\end{subfigure}%
\begin{subfigure}{.5\linewidth}
  \centering
  \includegraphics[width=\linewidth]{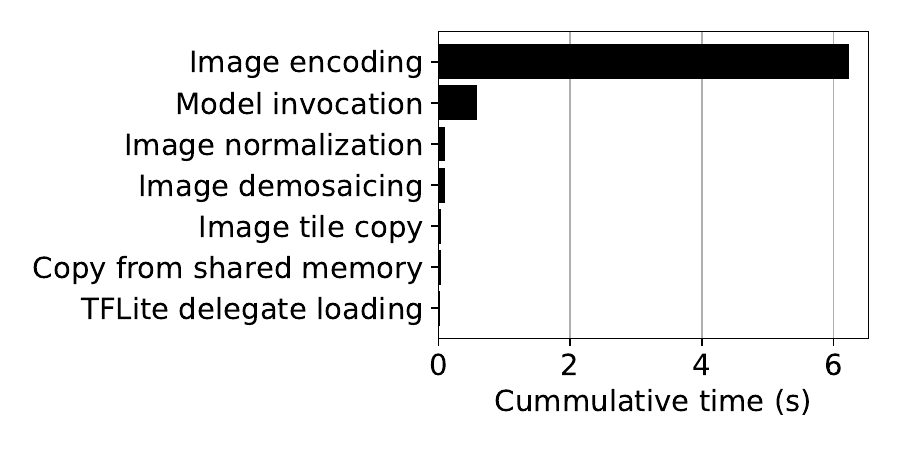}
  \caption{Merged (single-module) execution.}
  \label{fig:dipp-breakdown-single}
\end{subfigure}
\caption{Time spent on operations of an image processing pipeline. Only the operations that took more than $10$ milliseconds are shown. All invocations of the same operation in the pipeline are accumulated (e.g., memory copy).}
\label{fig:dipp-e2e}
\end{figure*}

\subsubsection{End-to-end pipeline}\label{sec:eval:dipp:e2e}

To evaluate the end-to-end performance of \DIPP, we compose a representative image processing pipeline with three modules.
The first module uses the OpenCV library to demosaic the raw BayerGR-formatted image into an RGB image,
which is then normalized.
The second module takes this RGB image and divides it into 224x224 pixel patches, which are then fed sequentially into a MobileNetV2 model \cite{mobilenetv2} for classification.
Fifth of these patches are then passed onto the last module, which uses the JPEGXL library \cite{jpegxl2021spec} to encode them. 

The input data used to evaluate this pipeline matches the expected format of images captured by one of the cameras to be deployed in the payload of \DISCO-2, 
2464x2056 pixels (5MP) images in BayerGR format with 12 bits per pixel.

To perform the machine learning task we leverage the TensorFlow Lite library \cite{tensorflow} together with its delegates, which can target the on-board NPU (\Cref{fig:sat-arch}) for the model invocation. 
MobileNetV2 was pre-trained on the ImageNet dataset \cite{imagenet} and fine-tuned for a land cover classification task on the RGB bands of the EuroSat dataset \cite{eurosat}. 
The size of the patches is adjusted to fit the model's expected input size and the ratio of the discarded images corresponds to 
a middle ground between estimated ratios of images discarded by cloud cover detection and more complex pipelines \cite{kodan}.

Finally, JPEGXL library is used to encode images before offloading them to a buffer, to be retrieved by a ground station. 
In this experiment we use the default settings of distance 1.0 and effort of 5, which correspond to visually lossless compression with a medium effort, providing a middle ground between the latency and the compression ratio.

To test the modularity and remote configuration of the pipeline, the end-to-end experiment also includes upload of the pipeline configuration and the configuration of the TensorFlow Lite module, parametrization of the model's file name, as would happen when the pipeline is first deployed.
However, the reported execution time does not include these steps as they happen once for the pipeline.

\Cref{fig:dipp-breakdown-multi} shows the cumulative time spent (sum across all operation invocations in a pipeline) performing the most expensive operations (amounting to at least 10 ms) of the pipeline, averaged over three runs. 
This figure shows that most of the time is spent on operations specific to image processing, such as encoding the images, model invocation, and image normalization and demosaicing, while little time is spent on auxiliary tasks, such as memory operations.
The average execution time of the entire pipeline for a single image amounted to $7.236$ seconds, most of which was spent encoding the final images.

\begin{table}[b]
\centering
\begin{tabular}{@{}llllll@{}}
\toprule
& \DIPP & demosaic.so & tflite.so & jpegxl.so & Static \\
\midrule
size KB & 1006 & 553 & 2003 & 405 & 3538 \\
\bottomrule
\end{tabular}
\caption{Size comparison for the pipeline (based on \Cref{sec:eval:dipp:e2e}) when dynamically loading modules vs linking the entire pipeline in the \DIPP~ binary (\textit{Static}).}
\label{tab:dipp-size}
\end{table}

\subsubsection{Impact of modular pipelines}\label{sec:eval:dipp:overhead}

To provide modularity and thereby save the uplink bandwidth using partial updates of pipelines, \DIPP~ promotes decomposition of the pipelines into separate modules.
\Cref{tab:dipp-size} shows the sizes of binaries needed to represent the image processing pipeline from \Cref{sec:eval:dipp:e2e}. 
By decomposing the pipeline into the separate shared object files, \DIPP~ achieves $43.4\%-88.6\%$ uplink savings in comparison to the statically linked version, which includes the end-to-end pipeline inside the \DIPP~ binary.

The pipeline decomposition furthermore allows \DIPP~ to execute each of these modules in isolated processes strengthening the robustness guarantees.
However, the forking of processes and the repetitive shared memory use to share data between processes can lead to overheads.
We therefore compare the execution of the pipeline described in \Cref{sec:eval:dipp:e2e} to its equivalent without decomposition, practically combining the three modules into one.
This means the entire pipeline is executed in a single process and requires shared memory access only in the beginning and at the end. 

\Cref{fig:dipp-breakdown-single} shows the cumulative time spent performing the most expensive operations (amounting to at least 10ms) of the single-module processing pipeline. 
Compared to \Cref{fig:dipp-breakdown-multi}, we can see that the shared memory operations add up to less time.
However, as suggested earlier in \Cref{sec:eval:dipp:e2e}, the time spent performing these operations is negligible in comparison to work on images performed by the pipeline, such as image encoding or machine learning inference.
The total time to execute this pipeline is $7.026$ seconds, which is $210$ milliseconds faster than the decomposed execution.
We find this overhead to be acceptable for the benefit it brings with respect to bandwidth savings and increased robustness.

\begin{figure}[t]
\centering
\includegraphics[width=\linewidth]{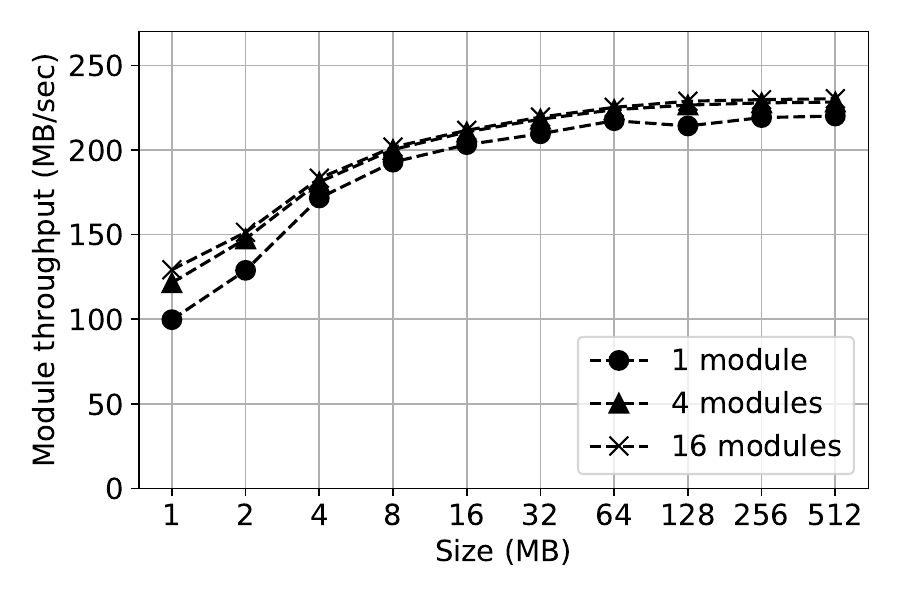}
\caption{Per-module throughput of pipelines with increasing number of modules and input data size.}
\label{fig:dipp-idt-throughput}
\end{figure}

\Cref{fig:dipp-idt-throughput} shows 
how per-module throughput is impacted by the number of modules in a pipeline and the input batch size. 
In this experiment, each module is an identity function, which copies the input batch data from shared memory into the local address space and back to shared memory right after.
%
We see that the most significant difference in throughput is with the lowest batch sizes, when the time spent inside the module is minimal.
However, the differences shrink with 
increasing size of the input data, as modules perform more work,
especially between the 4 and 16 module pipelines.
\DISCO-2's expected input batch size is at least $7.25$MB, which represents the size of a single raw image coming from the camera used in the end-to-end experiment.
Furthermore, this experiment represents an extreme case for the overheads with modules not performing any useful work.
Thus, the overheads would get amortized even further in real image processing pipelines.

\subsubsection{Robustness}\label{sec:eval:dipp:robustness}

To evaluate \DIPP's robustness against errors introduced by uploading faulty modules, we designed an adversarial module with various common errors:
division by zero,
dereferencing a null pointer,
accessing protected memory,
writing to read-only memory,
infinite recursion,
copying to invalid destination,
out-of-bounds,
calling a null function pointer,
allocating more memory than available,
freeing memory twice,
and buffer overflow.
\DIPP~ was able to gracefully recover from all of these errors, which cause the child process used for the adversarial module to crash.

\begin{figure}[t]
\centering
\includegraphics[width=\linewidth]{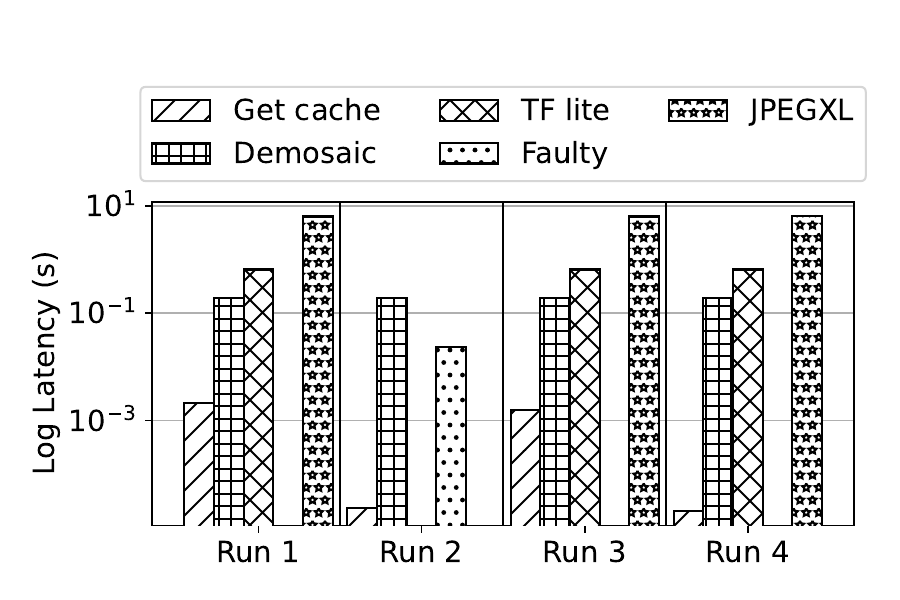}
\caption{Time spent executing pipeline modules. Runs 1, 3, and 4 execute a pipeline successfully, while run 2 executes a pipeline with a fault injected into its second module. Also shown is time to re-/build the cache in runs 1 (first invocation) and 3 (cache invalidated due to crash), and to reuse the cache in runs 2 and 4.}
\label{fig:dipp-robustness}
\end{figure}

To visualize this recovery behavior, we designed an experiment, executing two pipelines in succession. 
The first is the pipeline used in the end-to-end experiment in \Cref{sec:eval:dipp:e2e} (successful pipeline).
The second is the same pipeline, however, we replace the TensorFlow Lite module with the faulty module (faulty pipeline), crashing the pipeline execution by dereferencing a null pointer.
\Cref{fig:dipp-robustness} shows the results of this experiment.
It shows the execution latency of the pipeline modules, as well as, retrieving the pipeline configuration cache.
We first execute the first run of the successful pipeline.
As this was the first pipeline to be executed, building of the pipeline cache is triggered, leading to increased cache retrieval latency.
The second run executes the faulty pipeline, which reuses the cache built in the first run.
The JPEGXL module is never executed after the faulty module crashes, as the pipeline halts its execution and invalidates the cache to recover \DIPP~ into a known, safe state.
The third run executes the successful pipeline, rebuilding the invalidated cache, 
and the fourth run executes the successful pipeline again, leading to faster execution thanks to cache reuse.

\subsubsection{Memory footprint}\label{sec:eval:dipp:scalability}

\begin{figure}[t]
\centering
\includegraphics[width=\linewidth]{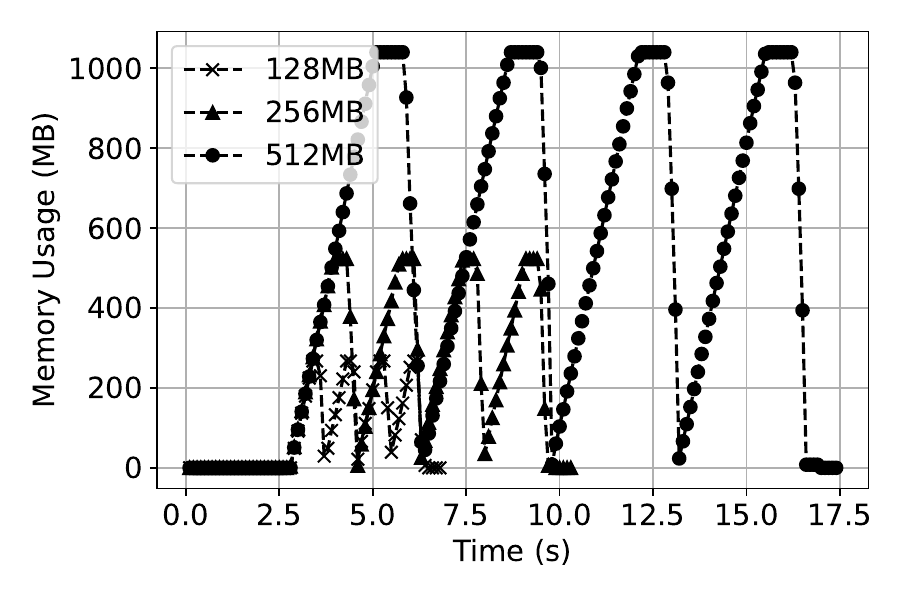}
\caption{Memory footprint of \DIPP~ processes during execution of a four module pipeline with increasing batch sizes. Each module performs identity operation (copies data from shared memory and back). The memory footprint increases up to $\sim 2\cdot size$, due to attached shared memory and the local copy.}
\label{fig:dipp-idt-mem}
\end{figure}

To analyze the memory impact of \DIPP~ 
we devise an experiment reusing the identity modules from \Cref{sec:eval:dipp:overhead}.
\Cref{fig:dipp-idt-mem} shows the memory usage of three four-module pipelines with varying input sizes.
We can see that the memory usage follows a predictable pattern, increasing to $\sim 2 \cdot$ image batch size and drops down to $\sim 0 $MB after the module finishes its execution, which holds for all pipeline executions.
This memory usage however only displays the memory usage of \DIPP~ and its child processes, and therefore disregards the size of the shared memory when none of the \DIPP~ processes are attached to it.
The system-wide memory usage of \DIPP~ therefore has the lower bound of the image batch size rather than $0$MB, until \DIPP~ removes the shared memory segment after its execution is done.

\subsection{\proc}\label{sec:eval:proc}
To evaluate performance of \proc, we first profile its memory utilization in \Cref{sec:eval:proc:memory}, followed by comparisons to the Lua scripting language in terms of the 
number of instructions in \Cref{sec:eval:proc:calls} and program sizes in \Cref{sec:eval:proc:serialization}.

\subsubsection{Memory usage}\label{sec:eval:proc:memory}

As the Cortex-M7 is highly limited in memory, careful memory management was one of the priorities when designing \proc.
After deploying \proc~and the CSP's networking stack, the data tightly coupled memory on the Cortex-M7 leaves us with $61,472$ B, which can be dedicated to the heap.
Main use of the heap on this core is the scheduler and its concurrently running procedures. 
We therefore profiled the memory usage of procedures (\Cref{sec:proc:procedure}) to determine the upper bound for the number of concurrent procedures and number of instructions in them.

The largest portion of heap for each procedure execution is the data structure holding the array of instructions. 
After this structure is allocated, the memory consumption is stable.
An exception to the constant memory consumption is procedure execution with non-tail nested call instructions.
The heap usage increase in this case is slow but has no upper bound.
We therefore provide a parameter to limit number of nested calls to prevent running out of memory.

\subsubsection{Call graph profiling}\label{sec:eval:proc:calls}

Since Lua cannot be deployed on the Cortex-M7 core as is, we have to extrapolate the performance of the two languages using their intermediate representation (IR).
We use the instruction count as a proxy for comparing their speed.


\begin{table}[t]
\centering
\begin{tabular}{@{}p{2.8em}R{2.8em}R{3em}R{3.8em}R{2.8em}R{4em}@{}}
\toprule
 & \proc & Lua & Lua (opt) & \shortstack[r]{Lua\\(golfed)} & \shortstack[r]{Lua (opt. \\+ golfed)} \\
\midrule
\# instr. & 293,096 & 358,210 & 11,771 & - & - \\
bytes & 114 & 712 & 290 & 252 & 108 \\
\bottomrule
\end{tabular}
\caption{Instruction count and program size of the intermediate representation of a program to find $10^{th}$ Fibonacci number in \proc, Lua with CSP state management, and optimized Lua (using local variables and reusing retrieved parameters). Golfing reduces whitespace and shortens variable names.}
\label{tab:proc-instructions-read}
\end{table}


We evaluate the two languages on a simple example program calculating the $n^{th}$ Fibonacci number.
\Cref{tab:proc-instructions-read} shows the number of read instructions, when the program is executed with $n=10$.
We can see that the performance of \proc~ and \textit{Lua} is similar,
which both have their
state fully encapsulated in CSP parameter table.
The \textit{optimized Lua} program relaxes this assumption and utilizes the CSP parameter table to retrieve values at the beginning of the execution and uses local variables for the rest of the execution.

From the results, it is obvious that encapsulating the state in CSP parameter table provides significant overhead.
However, encapsulating the state in CSP parameter table, provides a big advantage by allowing the scheduler to offload its state across the entire satellite's network, freeing up its highly limited heap for the execution of the concurrent procedures.

\subsubsection{Serialization}\label{sec:eval:proc:serialization}

As the procedures are to be uplinked to satellite using a network with highly-limited bandwidth, it is important to ensure their efficient serialization.
\Cref{tab:proc-instructions-read} provides a comparison of serialized size of various Fibonacci implementations.
The results show that \proc~ provides a very competitive result in comparison to \textit{Lua}, even when using the optimized version 
and golfing the implementation, i.e. removing all unnecessary whitespace and shortening variable names at the expense of readability. 


\section{Discussion}\label{sec:discussion}
\textbf{Applicability outside nanosatellites.}
While the development of \DIPP~ was motivated by the challenges of nanosatellites, it is not bound to this specific use case, and is especially suitable for remote deployments with poor connectivity such as marine use cases.
The \DISCO-2 camera control can be replaced by any other data source as long as it adheres to the communication interface \Cref{sec:dipp:comm} outlined.


\textbf{\proc's state management.}
\Cref{sec:eval:proc} showed pronounced differences between 
always relying on the CSP parameter table for the state and
utilizing on local variables after initial retrieval of the state over the satellite's network.
%
While we argue that keeping the state only in the parameter table reduces memory usage of the node running \proc, it could be beneficial to dedicate a portion of this memory to state to mimic the \textit{optimized Lua} implementation, especially in procedures that require tight timing.

\textbf{Tight CSP integration.}
As CSP provides a flight-tested software stack for satellite communication,
both \DIPP~ and \proc~ are tightly integrated with CSP.
However, \DIPP~ relies on CSP only for communication of the pipeline and module configuration and module shared object files.
This functionality can be easily achieved using other networking stack, as data transfer is a common feature.
\proc~ relies on CSP and its parameter table heavily, using it to encapsulate its state and distribute it over the nodes on the network.
We are not aware of other networking stack that provides a similar functionality out-of-the-box, and therefore the tight integration with CSP cannot be easily broken for \proc,
except for possible local reuse of parameters as mentioned above.


\textbf{Power efficiency.} 
\proc~ leverages the heterogeneous multiprocessing capabilities present on modern mobile platforms. 
The scheduler runs on a low-power real-time core and only wakes up the more powerful application cores, when needed for observations or processing. 
This approach significantly reduces baseline power consumption of the payload, while increasing the efficiency of the system under load.
Furthermore, since \proc~ has access to live telemetry data on the satellite, it provides operators the possibility of designing power-aware plans.
This, however, puts unnecessary burden on the operators and the uplink bandwidth. 
For future work, we would therefore like to make the scheduler aware of the both current power levels as well as predict future power generation of the satellite based on its orbit, which would serve as an admission criteria for the flight plans.

\section{Conclusion}\label{sec:conclusion}
This paper presented \DIPP~ and \proc, two novel systems designed to address the challenges of image processing on CubeSats.
\DIPP, a modular and configurable image processing pipeline framework, enables adaptation to evolving mission goals after deployment while maintaining robustness. 
\proc, a domain-specific language and runtime, allows for scheduling complex imaging workloads on resource-constrained processors found on CubeSats.

We demonstrate that \DIPP's pipeline decomposition adds negligible overhead while significantly reducing the network bandwidth required for updating pipelines. 
\DIPP~ also successfully recovers from various injected faults, which underlines its robustness.
\proc~ exhibits comparable expressiveness to Lua while achieving a good trade-off between memory requirements and performance.


\bibliographystyle{ACM-Reference-Format}
\bibliography{sample-base}

\end{document}